\documentclass[11pt, a4paper]{article}

\usepackage{url}
\usepackage{enumitem}
\usepackage{setspace}
\usepackage[margin=1in]{geometry}
\usepackage{graphicx}

\title{Preliminary Analysis of Potential Harms in the Luca Tracing System}
\author{Theresa Stadler$^1$ \and Wouter Lueks$^1$ \and Katharina Kohls$^2$ \and Carmela Troncoso$^1$}
\date{
    $^1$EPFL, \texttt{firstname.lastname@epfl.ch}\\%
    $^2$Radboud University, \texttt{kkohls@cs.ru.nl}\\[2ex]%
    \today
}

\begin{document}

\maketitle

In this document, we analyse the \textit{potential harms} a large-scale deployment of the Luca system\footnote{See \url{https://www.luca-app.de/system/}} might cause to individuals, venues,  and communities. The Luca system is a digital presence tracing system designed to provide health departments with the contact information necessary to alert individuals who have visited a location at the same time as a SARS-CoV-2-positive person. Multiple regional health departments in Germany have announced their plans to deploy the Luca system for the purpose of presence tracing. The system’s developers suggest its use across various types of venues: from bars and restaurants to public and private events, such religious or political gatherings, weddings, and birthday parties. Recently, an extension to include schools and other educational facilities was discussed in public.\footnote{See \url{https://twitter.com/_lucaApp/status/1368544491506896896?s=20}}

Our analysis of the potential harms of the system is based on the publicly available Luca Security Concept\footnote{The analysis is based on the March 10 version of the Security Concepts document (Document revision: eece55c available at \url{https://github.com/lucaapp/security-concept})} which describes the system’s security architecture and its planned protection mechanisms. The Security Concept furthermore provides a set of claims about the system’s security and privacy properties. Besides an analysis of harms, our analysis includes a validation of these claims.  

\newpage
\section*{Summary of Findings}
Our analysis of the Luca Security Concept uncovers the following concerns:\\

\noindent\textbf{Sensitive information leaked to the Luca Backend Server}. In its normal mode of operation, the Luca Backend Server collects and processes a large amount of sensitive information about venues and individuals. By design, the Luca Backend Server can:
\begin{itemize}
    \item Learn real-time information about venues, e.g., how many people are at a venue, when they arrive and when they leave. This information could be repurposed for, amongst others, commercial reasons or the surveillance of communities tightly linked to these venues, e.g., based on their political or religious orientation. 
    
    \item Learn which venues have been visited by Sars-CoV-2-positive users. This information could be repurposed to create a risk-based rank of venues. This might result in the social stigmatisation of venues, as well as their associated communities.  
    
    \item Learn the pseudonym of users who report a positive diagnosis for Sars-CoV-2, link the past location visits of these users for the period in which they were thought to be contagious, and learn the pseudonym of users that visited a venue at the same time as a positive index case. 
    These pseudonyms are unique, persistent identifiers of users and can potentially be linked to the user's IP address or phone number, which may lead to the re-identification of users.
    This information may be repurposed to take restrictive measures against individuals, threaten their freedom of movement and association, force users to change their behaviour, or contribute to users' social stigmatisation.
    
    \item Link visit records of the same (group of) users based on meta-data with high probability. This information could be repurposed for commercial interests or for the surveillance of individuals and communities.
\end{itemize}

These inferences do not require the adversary to actively modify the system’s operational information flows or circumvent its protection mechanisms. The Luca service operator (or any entity that compromises, coerces, or subpoenas the Luca Backend Server) can cause any of these harms without being detected. The risk of these harms may discourage the participation of users and venues, effectively reducing the effectiveness of the Luca system.\\

\noindent\textbf{High risk of abuse due to centralisation of trust}. The Luca system centralises trust in a single powerful entity, the party operating the Luca Backend Server. If this central entity acts maliciously, or is compromised, or coerced, the server could gain full access to any individual user's contact data and location history. 

The current system does not provide any technical safeguard against arbitrary access to this information in case of misbehaviour. Its security concept solely relies on procedural controls and requires full trust in the Luca service operator to follow the protocols faithfully.
\newpage
\section{Overview of the Luca System}
We give a high-level overview of the Luca system to support the analysis provided in this document. We only describe system aspects and information flows relevant to the analysis. We omit cryptographic details where they are not relevant for our analysis.\footnote{The full description of the Luca Security Concept is accessible at \url{https://luca-app.de/securityconcept/intro/landing.html}}

We note that Luca’s Security Concept does not explicitly describe the system deployment. At the time of this analysis, the source code of the Luca system is not openly accessible. Our description is a best-effort interpretation of the (sometimes implicit) clues provided by the Luca documentation. We cannot be sure, however, whether the actual implementation follows these exact interaction patterns. In particular, we do not know whether the actual implementation includes backend servers not mentioned in the documentation. We note that any differences between our assumptions about the system deployment described below and the actual deployment scenario might affect the harm analysis put forward in this document. To avoid overestimating the harms, our assumptions about implementation take the least harmful option or explicitly explain the different possible implementation choices.

\begin{figure}
    \centering
    \includegraphics[width=\textwidth, trim=0 2cm 0 0]{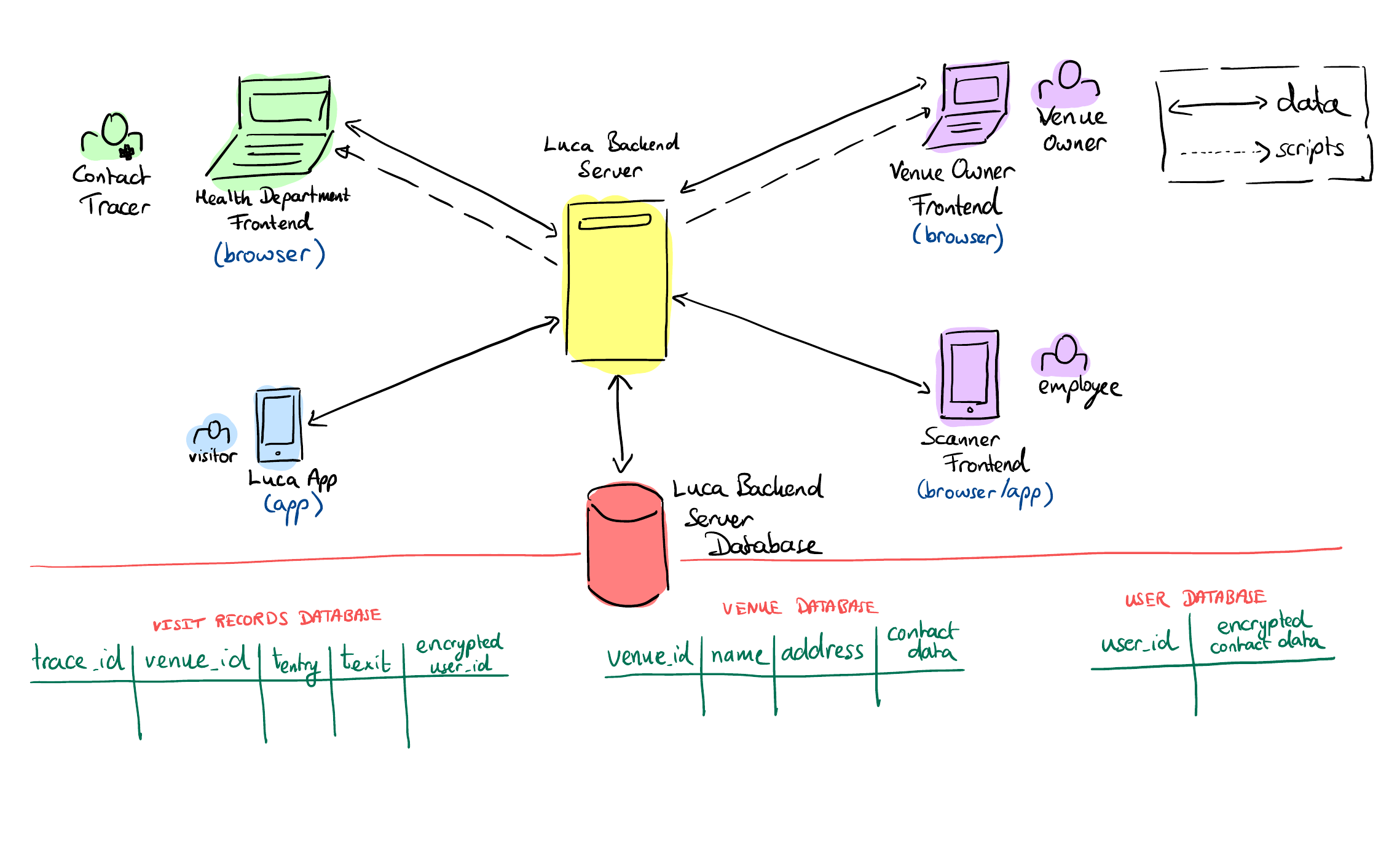}
    \caption{Overview of the Luca system}
    \label{fig:luca-overview}
\end{figure}

At the centre of the system is the Luca Backend Server. This server stores a record of every visit made by a visitor to a Luca-supported venue provides an interface to Health Departments and venues, and orchestrates the tracing process. The Luca Backend Server also holds a database of registered venues and a database of users linked to a permanent pseudonymous identifier, the \texttt{user\_id}. This pseudonym is linked to the user’s encrypted contact data, and during the system’s operation, it can be associated with metadata as we detail in our analysis. See also Figure~\ref{fig:luca-overview}.

\noindent\textit{Interacting with the system}. In our analysis, we assume that Health Departments, venues, and visitors use the following mechanisms to interact with the Luca system:
\begin{itemize}
\item Health departments use the Health Department Frontend. This is a web application that interacts with the Luca Backend Server. Health departments load this web application (e.g., the corresponding HTML and Javascript files) from the Luca Backend Server.
\item Venues Owners use the Venue Owner Frontend to interact with Luca. Again, this is a web application that interacts with the Luca Backend Server. Venue Owners load this web application from the Luca Backend Server.
\item Venue employees use the Scanner Frontend to register visitors. This frontend is either a web application (as above) or a smartphone/tablet application.
\item Visitors use the Luca App, a mobile application installed on a visitor’s personal device, to interact with the Luca Backend Server.\footnote{The Luca system provides the option to record location visits using a static Badge that is scanned by the Venue or a web application. For simplicity, we only describe the flows using the Luca app, as we expect that the are more likely to be used.}
\end{itemize}

\subsection{Joining the system}
\textit{Users} join the system using the Luca App installed on their personal device. The server validates the user’s phone number and stores an encrypted record of the user’s contact data (name, address, phone number) at the Luca server. The phone receives an associated \texttt{user\_id}. The user’s device holds the secret key that unlocks the encrypted contact record. The phone also generates a tracing seed that is refreshed daily.

\textit{Venues} join the system by creating an account at the Luca Backend Server. Each venue enters their information (name, address, etc.) into the Venue Owner Frontend, and the Luca Backend Server stores the venue record in a central database. The Venue Owner Frontend generates a public-private venue keypair. Venues store the private key locally. The public keys of all venue owners are stored at the Luca Backend Server to facilitate QR code check ins.

\textit{Health departments} join the system by requesting the Luca Backend Server to create a new Health Department account. The Luca service operator provisions a certificate and account credentials for the employees of the Health Department. Employees use their account credentials and certificates to access the Health Department Frontend. For each newly registered Health Department, the Health Department Frontend generates a new Health Department key pair. The private key of each key pair is stored locally at the responsible Health Department.

To facilitate the decryption of contact data in visit records, Health Departments share daily master keys. Every day, Health Departments use an asynchronous process to generate a new shared master key. We detail this process in Section~\ref{sec:active}.

\subsection{Visiting a venue}\label{sec:overview:visit}
For each new check-in, the Luca App derives a new \texttt{trace\_id} from its tracing seed. The Luca App also obtains the current day’s public master key from the Luca Backend Server. It creates an encrypted user reference by encrypting the user’s \texttt{user\_id} and contact data secret key against this master public key.
The Luca App displays both the \texttt{trace\_id} and the encrypted user reference in a QR code. This code is scanned by the venue’s Scanner Frontend. The Scanner Frontend adds a second layer of encryption to the user reference using the venue’s public key. It uploads a check-in record that contains a unique identifier for the scanner, the user’s current \texttt{trace\_id}, the double-encrypted user reference, and the check-in time to the Luca Backend Server.
The Luca App polls the server with the current \texttt{trace\_id} to confirm that the check-in was successfully recorded.
Once the user leaves the venue and triggers check-out, the phone sends the \texttt{trace\_id} of the current check-in together with the departure time to the Luca Backend Server.

The Luca proposal also foresees two other check-in modes: self check-in, where visitors use their app to scan a QR code provided by the venue, and badge check-in, where venues scan a static QR code held by the visitor. These flows’ privacy properties are the same or worse than those of the flow sketched above. Hence, they can only worsen the resulting harms. We will note the differences where relevant.

\subsection{Tracing and notification}
Tracing proceeds in two phases. An infected user can initiate tracing by encrypting its \texttt{user\_id} and the tracing seeds for the relevant time period to the current daily master key and uploading them to the Luca Backend Server. The responsible Health Department logs into the Health Department Frontend and retrieves from the Luca Backend Server the encrypted \texttt{user\_id} and seeds as well as the daily private master key encrypted to their specific Health Department key. The Health Department Frontend unlocks the daily master key and uses it to decrypt the user’s identifier \texttt{user\_id} and tracing seeds. It sends the decrypted identifier and tracing seeds back to the Luca Backend Server. The server uses these seeds to find all check-in records associated with the positive index case via their \texttt{trace\_id}s. This allows the Luca Backend Server to identify which venues the positive index case visited, when, and for how long.

To provide the contact data of users that visited a venue at the same time as the index case, the Luca server proceeds as follows. It contacts the venue owner to request their assistance. The venue owner logs in to the Venue Owner Frontend and provides their private key. The Venue Owner Frontend requests the doubly encrypted records of the Luca Backend Server's relevant times, decrypts them locally, and sends the singly encrypted records back to the server.

The Health Department Frontend retrieves the singly encrypted records from the Luca Backend Server and decrypts them with the corresponding daily master key to obtain the \texttt{user\_id} of all relevant users. Finally, the Health Department Frontend requests the encrypted contact records corresponding to the identified \texttt{user\_id}s and decrypts them.

\section{Analysis of Potential Harms}
We analyse the system with respect to the potential harms its deployment might cause to individuals, venues, and communities. We evaluate whether the system defends against each of these harms and, where applicable, discuss whether it mitigates these harms through architectural design decisions, through technological safeguards, or through procedural controls.

\subsection{Potential Harms}
We have identified the following harms as the most concerning and relevant for our analysis:\\

\noindent\textbf{Targeted surveillance of individuals}\\
An adversary might re-purpose the system to extract information about individual users, such as personal contact data, present and past location visits, or social relationships. If the system's deployment leaks this information, it might lead to the surveillance of individuals and the associated restrictions of individuals' freedoms. 

\noindent\textbf{Targeted surveillance of venues and  associated communities}\\
An adversary might re-purpose the system to extract information about venues and their activities. This includes, for instance, the location of a registered venue itself, information about the number of people assembling at a specific location and time, or any metadata associated with a venue, such as the name and contact data of the responsible venue owner (or event organizer). Registered venues might include public and private events, such as religious gatherings, political events, or family celebrations which renders this information highly sensitive.
If the system's deployment leaks this information, it might lead to the surveillance of target groups based on venues these groups are likely to visit. 

\noindent\textbf{Social stigmatisation of individuals}\\
An adversary could use the system to learn which individuals have reported a positive test for Sars-CoV-2 and which individuals the responsible Health Department might notify about a potential infection. Leakage of this information to any entity other than the responsible Health Department is a clear violation of user privacy, and can have further consequences. If this information were to be made public, it could lead to the social stigmatisation of individuals. More worryingly, it could lead to the coercion of individuals. Adversaries could threaten to publish compromising information about individuals to exert pressure. This is particularly problematic for public persons whose reputation might be harmed by revelations about their health status or their presence at particular events. Target individuals could be forced into changing their public behaviour which can have negative effects for society at large.           

\noindent\textbf{Social stigmatisation of venues and their associated communities}\\
An adversary could use the system to learn which venues have been visited by a positive index case. If this information were to be made public, it might lead to the social stigmatisation of venues visited by a high number of positive index cases. This could further cause harm to the communities tightly linked to these venues based on, for example, their political, sexual, or religious orientation.

\subsection{Analysis of Potential Harms in the Luca System}
A deployment of the Luca system might cause the following harms to individuals, venues, and communities.\\

\subsubsection{Targeted surveillance of individuals through the Luca Backend Server}
\label{sec:surveil:users}

\noindent\textit{User location profiles}. Whenever a user checks in to a venue or departs from a venue, their personal device makes a direct connection to the Luca Backend Server and provides the user’s current \texttt{trace\_id} (see Section \ref{sec:overview:visit}). For each connection, the Luca server observes the user’s IP address and other information such as the user’s device type and possibly the Luca App version. The Luca Backend Server can use this metadata to probabilistically link check-in records that likely belong to the same user and learn this user's (partial) location history.

For users with a unique IP addresses -- e.g., every time mobile phones are assigned an IPv6 address, and exceptionally when the gateway does not use an IPv4 NAT -- the Luca Backend Server can link check-in records to the same user for as long as the user's IP address does not change. The server first observes a visit record from the venue associated to a \texttt{trace\_id}, and then a poll request associating an IP to this \texttt{trace\_id}. Thus, the server can associate IPs to visit records; and can link records over time via this IP.

In most cases, mobile providers deploy a carrier-grade NAT to share IPv4 addresses among users.\footnote{This NAT-based sharing is mentioned as mitigation against linkability risks in the Luca Security Concept: \url{https://luca-app.de/securityconcept/processes/guest_app_checkin.html}} 
As a result, the Luca Backend Server observes the same IPv4 address for several devices. Even in this scenario, however, the Luca Backend Server is likely to be able to link a user's check-ins across locations and corresponding \texttt{trace\_id}s. To do so, an adversary with access to the Luca Backend Server can leverage the following observations to reduce the anonymity set of users behind the same public IPv4 address:

\begin{itemize}[nosep]
    \item Devices with different mobile phone carriers will always be distinguishable based on their IP address, as it refers to an operator's gateway.
    \item Even when using carrier-grade NAT, mobile devices likely maintain the same IPv4 address over a prolonged time period. For technical reasons, the number of devices sharing the same external IPv4 address is limited.
    \item Of the devices sharing the same IPv4 address, a significant fraction might not be using the Luca App.
    \item The remaining devices sharing an IPv4 address are likely not all in close physical proximity.
    \item The check-in events recorded by venues contain additional metadata such as the user's device type\footnote{See Luca Security Concept: \url{https://luca-app.de/securityconcept/processes/guest_app_checkin.html}} which further reduces the likelihood that two devices of the same type share an IP address.
\end{itemize}
As a result, the user anonymity set is likely to not be large enough to ensure privacy.

To validate our claims, we run a preliminary analysis in which we use a phone to make requests to our own HTTP server. We use the server's access log to analyze the visiting IPv4 and IPv6 addresses and the ports used in the case of IPv4. At the same time, we record the phone's traffic to compare the internal and external public addresses.

This small experiment shows that as long as the user does not disconnect from the network (e.g., restarting the phone, changing the SIM card, or switching on airplane mode), the IP address is stable. Furthermore, we see that each new IPv4 request at an HTTP server uses an individual port and that these ports are incremented over time. When visiting multiple locations, users might be tracked through these port numbers even though their external IP address is behind a NAT. Moreover, we see that carrier-grade NAT is only used for IPv4 addresses. If carriers assign IPv6 addresses to devices (e.g., in T-Mobile and Vodafone) and the Luca server is reachable via IPv6, the server can directly observe the client's unique IPv6 address.

In a second experiment, we analyze the geographical distribution of gateways within Germany. This provides us with information about the gateways users share and the anonymity sets that the public IPv4 addresses of these gateways represent. We observe that the gateway that users are assigned to does not depend on their geographical location.

While our experiments are limited in scale and require validation at a larger scale, our results demonstrate that an in-depth study of mobile and WiFi network behaviour is needed to substantiate the claim that linkability is not possible. We also note that increasing the level of unlinkability may not depend on the service provider (in this case, the Luca server) but on the carrier. In fact, it may be that the only way to guaranteeing unlinkability is to rely on anonymous communications systems.

We conclude that the network-level meta-data already enables the Luca server to drastically reduce the anonymity set of check-in records, eventually enabling probabilistic linking of different checks. To increase the strength of this inference, the Luca Backend Server can combine this network-level data with the location of venues and check-in and check-out times, which it can also observe. Combining the IP-based analysis with a spatiotemporal analysis based on check-in and check-out times and the location of venues enables the Luca Backend Server to construct (partial) location profiles linked to a pseudonymous user identifier. \\

\noindent\textit{Social relationships}. In addition, the Luca server can probabilistically link records that belong to groups of users. Users in a group that arrive at a venue together are bound to check-in within a short time window using the same scanner. For each check-in, the Luca server sees precise check-in times and the \texttt{scanner\_id} for each record. Moreover, a group of users likely departs at the same time which again results in a set of check-out events within a short time window. The Luca Backend Server can combine these data to probabilistically group records by user groups.

This enables the Luca Backend Server to reconstruct relationships between records of different users. And as we explained above, records of individual users can be grouped by temporary pseudonyms, thus enabling the Luca Backend Server to recover pseudonymous relationships. However, should these pseudonyms become linked to an identifier (e.g., through their IPs or other meta-data; or as a result of tracing) then the server can reconstruct meetings between individuals and their duration. This surveillance harms users and communities.\\

\noindent\textit{Active surveillance}. The probabilistic linkage attacks described above assume that the Luca Backend Server does not actively try to circumvent the confidentiality provided by the double encryption of check-in records. As we detail in Section~\ref{sec:active}, however, several methods enable an active adversary with control over the Luca Backend Server to do so. Using these methods, the adversary can reveal the \texttt{user\_id} that belongs to any check-in record of its choice. This enables the adversary to create location traces linked to a permanent pseudo-identifier.\\

\noindent\textit{User re-identification.} A malicious backend server can leverage other information about users, such as a user’s phone number used during registration or their IP address, to attempt to re-identify individuals. If the Luca Backend Server successfully matches a user's pseudo-identifier to an identity, any information the server holds about this pseudo-identifier, such as (partial) location traces or a user's health status, becomes directly linked to the re-identified individual.\\

\noindent\textit{Surveillance of individuals.} All of the inferences described above violate users' location privacy and lead to the surveillance of individuals. The powers of the central server could easily be abused to observe the whereabouts of target individuals or to reconstruct social relationships between individuals.

The current system design does not include any technical safeguards against many of these adversarial inferences. For instance, an adversary can conduct probabilistic linkage attacks based on user IP addresses without any changes to the system's normal operations. Confidentiality of user's location traces relies solely on the trustworthiness of the Luca Backend Server. If the Luca Backend Server acts maliciously, is compromised, or coerced to provide access, it can cause harm without being detected.

\subsubsection{Surveillance of venues and their associated communities through the Luca Backend Server}
\label{sec:surveil:venues}

\noindent\textit{Electronic registry of events}. The system design requires venues to register at the Luca Backend Server. The server stores the provided venue information in a central database, including the owner’s or organizer's contact information and the exact geo-coordinates of the venue. Such a database allows any entity with access to the Luca Backend Server to learn about the existence of these venues.

For some types of venues, such as political or religious gatherings, creating such records may in itself pose a threat. While many venues, such as bars or restaurants, are already part of publicly available registries, e.g., for licensing or mapping purposes, for other types of social gatherings relevant in the context of contact tracing, no electronic records exist. Creating a digital record of these events, stored in a central database, might cause harm to communities and restrict their fundamental rights to freedom of association.\\

\noindent\textit{Real-time profiling of venues.} During normal operation, the server collects a record of every check-in event reported by venues together with a unique identifier for the scanner that recorded the check-in. The \texttt{scanner\_id} enables the server to link a check-in back to a specific venue. Check-ins are sent to the Luca Backend Server in real-time to allow users’ devices to confirm with the server that a scan was successful. Later, users directly communicate their check-out times to the central server. 
The Luca Backend Server hence observes, in real-time, how many people are gathered at a venue and when they arrive and leave. This information could be easily repurposed for the (real-time) surveillance of venues and their associated communities.

The system does not include any mitigations against this harm. The Luca Backend Server, by design, acts as a central entity that collects and links information about venues and their activities. The system does not aim to prevent the Luca Backend Server from accessing this information.
While the current system documentation does not mention any secondary use of the data collected at the Luca Backend Server, the Luca service operator could at any time decide to repurpose the data for commercial purposes, could be coerced to share this information, or compromised by unknown actors.

\subsubsection{Social stigmatisation of individuals}
\label{sec:stig:users}
\noindent\textit{Positive users.} To initiate the tracing process and share the relevant contact data with the responsible Health Department, individuals use the Luca App to upload an encrypted user reference, including their \texttt{user\_id} and tracing seeds, to the Luca Backend Server. The server issues a verification code that the user's app displays. The upload process allows the Luca Backend Server to link the verification code is issued to a user's IP address.\footnote{Users are likely to initiate the tracing process while quarantining at home so that an upload can be linked to a user's home IP address.}\\

The Health Department uses the user-specific verification code to retrieve the encrypted user reference of the positive index case from the Luca Backend Server. It decrypts the user reference to obtain the user's \texttt{user\_id} and requests from the Luca Backend Server the encrypted contact data stored for this \texttt{user\_id}.
The Luca Backend Server hence observes, within a short time frame, two requests from the Health Department: first, a request for the user reference linked to a verification code, and second, a request for the contact data linked to a \texttt{user\_id}. The server can correlate these requests to link user pseudonyms to verification codes and their corresponding IP addresses.\footnote{This risk is explicitly acknowledged in the Luca Security Concept: \url{https://luca-app.de/securityconcept/processes/tracing_access_to_history.html}}
      
To identify the venues visited by a positive index case, the Health Department forwards the \texttt{user\_id} and the corresponding tracing seeds to the Luca Backend Server who identifies all past location visits recorded for this user. 
Through the tracing process, the Luca Backend Server learns which \texttt{trace\_id}s can be linked to the same \texttt{user\_id}. This reveals the past location visits of the positive index case to the Luca Backend Server.\\

In summary, the Luca Backend Server can observe the IP address of users who report a positive diagnosis and correlate this IP address to a permanent pseudo-identifier, the \texttt{user\_id}. Subsequently, the server learns the full location history linked to the pseudonymous user, including the geo-coordinates of all venues visited and exact arrival and departure times.\\

\noindent\textit{Trace contacts.} To obtain the contact data of traced individuals who need to be notified, the Health Department Frontend obtains a list of their \texttt{user\_id}s and shares them with the Luca Backend Server. This allows the Luca Backend Server to infer which (pseudonymous) users have been in contact with a positive index case.\\

\noindent\textit{Breach of user confidentiality.} The inferences described above clearly breach the confidentiality of user's sensitive health and location information. If made public, the learned information could lead to the stigmatisation of users who have either tested positive for Sars-CoV-2 or visited a venue at the same time as a positive index case. This risk could discourage users from reporting a positive test result or from participating in the system altogether.
Even more worryingly, adversaries could use the information to exert pressure on individuals by threatening to publish compromising information.\\  

The system does not provide technical safeguards to prevent these harms. While it ensures that the Luca Backend Server does not hold any cleartext records of user’s personal data, i.e., it only identifies users via a pseudonymous identifier. The system still allows the central server to link sensitive data, such as multiple check-ins and encrypted user reference uploads, to these pseudo-identifiers. The more data the server is able to link to the same user profile, the more likely it becomes that the user might be identifiable based on the associated metadata (see \ref{sec:surveil:users}, Re-identification).

\subsubsection{Social stigmatisation of venues and their associated communities}
\label{sec:stig:venues}
Health departments rely on the Luca Backend Server to identify which venues a positive index case has visited in the past. During the tracing process, the Health Department Frontend sends the tracing seeds of users who reported a positive test for Sars-CoV-2 to the Luca Backend Server. The server searches the database of check-in events for \texttt{trace\_id}s that belong to a tracing seed marked as positive. For this purpose, the server must identify which venues should be contacted and the relevant tracing times.

The Luca Backend Server not only learns which venues have been visited by a Sars-CoV-2-positive user, but also how many people were present at the same time as the positive index case, and when the incident occurred. This information could be used to rank venues based on their positive case number and result in social stigmatisation of venues and their associated communities. Social groups affiliated with specific venue types, such as religious or political gatherings, might suffer negative consequences from being perceived as “dangerous” or “negligent”. The potential harms of being publicly marked as a high-risk location could discourage venues’ participation in the tracing process.

The system does not include any mitigation against this harm. Due to its central role in the tracing process, the Luca Backend Server gains a detailed overview of epidemiologically relevant information that can be repurposed.
To avoid this leakage, the system design should ensure that only responsible Health Departments learn which venues have been visited by a positive index case, and for which time frames contact data needs to be requested. Given the current architecture, eliminating this information would entail a substantial redesign of the protocols and information flows.

\subsection{Active attacks}\label{sec:active}
The confidentiality of records stored at the Luca server rests on the double encryption provided by venues and Health Departments. An adversary who actively circumvents these protection mechanisms can learn the exact check-in history of users and their identities (e.g., who was where, when).

We now detail several methods which an adversary with control over the Luca Backend Server can use to defeat the system's confidentiality protections. The methods described below all imply that the Luca Backend Server deviates from its normal execution path and does not follow the policies laid out in the Luca Security Concept. This might be the case if the Luca service operator has a high incentive to act against these policies, or is coerced, subpoenaed, or compromised by outside actors. The analysis demonstrates that the Luca Backend Server is a fragile single-point-of-failure for the confidentiality of the Luca system.

\subsubsection{Circumventing the protection provided by venue encryption}
\label{sec:active:venue}
The Luca Backend Server has the following means to circumvent the protection offered by encrypting records against the venue’s public key.

\begin{enumerate}
    \item \textit{Directly request decryption from venues}. In the current design, Venue Owners cannot authenticate the origin of decryption requests as all requests are channeled via the Luca Backend Server. The Venue Owner merely acts as a decryption oracle, i.e., it cannot know whether the decryption of records is requested based on a legitimate tracing query from a Health Department or whether this is a malicious request from the Luca Backend Server. 
    \item \textit{Expand legitimate decryption requests}. The Luca Backend Server is responsible for forwarding and mediating legitimate decryption requests by the Health Department Frontend. The Luca Backend Server is expected only to request the decryption of records that fall within a time slot relevant for contact tracing.
    However, the Luca server can arbitrarily expand the requested time intervals, adding as many other records as it wants to this decryption request. The Venue Owner currently has no means to detect that these extra records are outside of the range authorized by the Health Department.
    \item \textit{Substitute a venue's public key}. Users have the option to check-in by scanning a printed QR code provided by the venue. This QR code currently does not include the venue’s public key. Instead, the app retrieves the venue's public key from the Luca Backend Server. The server can, therefore, trivially replace the correct key with an encryption key of its own. It can then trivially remove one layer of encryption.\footnote{Incorporating a venue's public key into the printed QR code is left as a future improvement to the system implementation: \url{https://luca-app.de/securityconcept/processes/guest_self_checkin.html}.}
    \item \textit{Silent modification of the Venue Owner Frontend code}. The Luca Backend Server provides the code that runs at the Venue Owner Frontend. As such, the server could modify the code directly or do so indirectly via one of the included JavaScript libraries. The Venue Owner Frontend could be modified to (a) generate backdoored venue keys; (b) to exfiltrate a copy of the private key once it is generated; (c) to exfiltrate a copy of the private key any time it is used; or (d) to circumvent any of the other checks around accepting decryption requests.\\
    This attack can be targeted. For example, the server can send modified code only to a venue for which it wants to obtain the decryption key. Such targeted attacks are thus extremely difficult to detect.
\end{enumerate}

\subsubsection{Circumventing the protection provided by Health Department encryption}
\label{sec:active:hd}

The Luca App uses a daily master public key, distributed via the Luca Backend Server, to encrypt the user's contact data before exposing it via a QR code to the Scanner Frontend. The system's confidentiality relies on the fact that only Health Departments can access the daily master private key to remove this inner layer of record encryption.
Every Health Department can access this daily master private key. We demonstrate why this shared secret is a major weakness of the Luca design.

\noindent\textit{Daily master key rotation}.
We first detail the process used to generate and rotate the daily master key pair. Every Health Department holds an encryption and a signing key pair. The Luca Security Concept does not specify how these keys are stored. We assume that all private keys are stored locally at the Health Department and entered into the Health Department Frontend when needed.

If a new daily master key pair must be generated, the first Health Department that logs in to the Health Department Frontend proceeds as follows:
\begin{enumerate}[nosep]
    \item Compute a new daily master key pair
    \item Sign the new master public key (using its Health Department's private signing key) and upload both the new public key and the signature to the Luca Backend Server.
    \item Retrieve the public encryption keys of all the other health authorities from the Luca Backend Server. Encrypt the master private key for each of the other Health Departments. Upload these ciphertexts to the Luca Backend Server.
\end{enumerate}
Whenever a Health Department needs a daily master private key, they request the encrypted master private key from the Luca Backend Server and decrypt it using their private decryption key.

In the current version of Luca, the public keys of Health Departments do not come with a certificate. We therefore analyze the system under the assumption that no key certification is in place. Adding certificates adds some protection to the system but does not protect against all attacks (see Section~\ref{sec:protect-active}).\\

\noindent\textit{Circumventing Health Department encryption}. The Luca Backend Server has the following means to circumvent the protection offered by encrypting records against the daily master key.

\begin{enumerate}
    \item \textit{Substitute the daily master public key}. The Luca App and the Scanner Frontend retrieve the daily master public key from the Luca Backend Server. For purposes of authentication and integrity, this key is signed by a Health Department’s public key. The Luca server also provides the signature and the corresponding public key to verify the signature. As long as there are no certificates in place to bind the signing key to a legitimate Health Department, the Luca Backend Server can substitute this daily master public key with a key of its choosing and create its own signature. Currently, this key substitution cannot be detected. After replacing the daily master key, the Luca Backend Server can decrypt any records encrypted under this key. This is particularly problematic as the daily master public key is also used to protect the confidentiality of users who report a positive diagnosis for Sars-CoV-2 (see Section \ref{sec:stig:users}). 

    \item \textit{Impersonate a Health Department}. The Luca Backend Server can impersonate a Health Department to learn the daily master private key. To do so, in step (3) of the daily master key generation protocol, it adds its own public key to the list of public keys of Health Departments. An honest Health Department cannot distinguish the Luca key from legitimate Health Department’s keys. Therefore, it will encrypt the new daily private key against the public key of the Luca server and return it to the Luca Backend Server. The Luca server can then decrypt it to obtain the daily private key. We note that in the the current implementation, the Luca Backend Server has the power to enroll any party as a Health Department. Proper use of certificates would prevent this attack.

     \item \textit{Using the Health Department Frontend as a decryption oracle}. A malicious Luca Server can use the Health Department Frontend as a decryption oracle to obtain the \texttt{user\_id} of (partially decrypted) check-in records. 

    During the tracing process, the Luca Backend Server reconstructs the visit history of positive users and identifies which check-in records need to be decrypted to obtain the contacts of users who should be notified. It sends the relevant records to the venues who remove the records' outer layer of encryption. The server forwards the partially decrypted records to the Health Department Frontend who removes the last layer of encryption and sends the resulting \texttt{user\_id}s back to the Luca server. The Health Department Frontend currently has no way to validate the authenticity of the decryption requests. A malicious Luca server can request the Health Department Frontend to decrypt any record of its choice.

    This attack on the system's confidentiality may be detected. If the responsible Health Department calls all users included in the list of records forwarded by the Luca Backend Server, the Health Department might learn that some of the notified users have not visited any relevant venue.\footnote{To avoid being caught, a malicious Luca Backend Server could decide not to return the contact data of records outside the tracing window. For instance, the implementation of the Health Department Frontend might silently drop such invalid records, or show a cryptic error message.}

    \item\textit{Silent modification of the Scanner Frontend code}. In one of the possible deployment modes, venues use a Scanner Frontend in the form of a web application. In this case, as above, the Luca server can modify the JavaScript code and subtly replace the daily master key with one of its own or circumvent certificate checks.

    Such an attack, especially when targeting only a few venues, is likely to go undetected.

    \item\textit{Silent modification of the Health Department Frontend code}. The Luca Backend Server provides the code that runs at the Health Department Frontend. As such, the Luca server could modify the code directly or do so indirectly via one of the included JavaScript libraries. As a result, the Luca server can modify the Health Department Frontend code to (a) generate backdoored Health Department encryption keys; (b) to exfiltrate a copy of the private encryption key once they are generated; (c) to exfiltrate a copy of either private key any time they are used used; or (d) to subtly circumvent certificate checks. Each of these modifications results in the Luca server learning all daily private keys.
 
    This attack can be targeted and therefore executed stealthily. Modifying the code for one single Health Department for one single session or page-load is sufficient to obtain the Health Department’s private decryption key. And thereby all past and future daily master private keys.

\end{enumerate}

In addition, there is a risk that any of the Health Department's private keys might be leaked. The large number of private keys issued\footnote{There are more than 400 Health Departments in Germany.} to access the shared master key considerably increases the risk that one of these keys might be leaked or stolen. Access to a Health Department private key would allow the Luca Backend Server to access all past and future daily master keys.

\subsubsection{Protections against active attacks}\label{sec:protect-active}
The attacks explained above demonstrate that the confidentiality of check-in records hinges on a small number of entities' trustworthiness. In particular, a maliciously acting Luca Backend Server can quickly compromise the entire system's security and confidentiality. Some of these weaknesses, however, can be addressed.  We classify the above attacks into three categories:

\noindent\textit{Attacks mitigated by certificate extensions}. Some attacks can be mitigated using well-known security mechanisms. As the Luca Security Concept mentions, embedding the venue’s public key directly into the printed QR code prevents substitution attacks. Extending the design with a trusted Public Key Infrastructure (PKI) would limit the remaining impersonation and key substitution attacks. The use of a PKI, however, requires careful implementation and only partially addresses the design's trust issues: The entity chosen as certificate authority must be a third party that has to be trusted not to collude with the Luca Backend Server, not to enroll any entity in the system that is not a Health Department, or to impersonate a Health Department itself. The certificate authority also must ensure that the Luca Backend Server cannot create its own valid certificates.

\noindent\textit{Attacks mitigated by cryptographic extensions}. These are decryption oracle attacks where the Luca server asks Venue Owners or Health Departments to decrypt records outside the relevant tracing windows. Such attacks could potentially be mitigated by cryptographic extensions of the current design. These changes, however, might, in and of themselves, open up new attack vectors.\footnote{As an example of how additional design complexity can lead to an increase in risk, consider the daily key rotation of the shared master key. While the daily key rotation might reduce the impact of the leakage of a single daily master private key, the process to generate a new master key pair actually opens up new ways to obtain access to the master private key.} Therefore, implementing additional cryptographic techniques has to be done carefully and should involve domain experts, and external and public reviews of system design and implementation.

\noindent\textit{Attacks inherent to the design}. Attacks that leverage the fact that all Health Departments share a master decryption key or that trusted code is provided by untrusted platforms are largely inherent to Luca’s design choices. Luca must likely be completely redesigned to avoid storing security-critical private keys at every Health Department. Moreover, in the current design, the code for Health Departments and venues is provided by a party that, according to Luca's security concept, is not trusted to keep users’ data confidential. The risk of misbehaviour is amplified because it is possible to modify code in subtle ways such that malicious modifications are hard, if not practically impossible, to detect. Finally, we note that such misbehaviour cannot be prevented by opening the source code, as this code could easily be modified under deployment requiring intensive oversight to be detected.

\section{Conclusions}
In this document, we have provided a preliminary analysis of the potential harms that might result from the reuse or abuse of the information collected and made accessible by the Luca system. 

\vspace{2mm}
\noindent\textbf{Main findings}. Our analysis demonstrates the following main concerns:

\begin{itemize}
    \item Through their interactions with the system, users and venues generate a large amount of sensitive information that, by design, is made accessible to the Luca service operator. If the Luca service operator acts maliciously, either spontaneously or under coercion, or is compromised, it can obtain further sensitive information about users. This might eventually allow the Luca service operator to track individual users across venues and reveal social relationships between users.
    
    \item The richness of the information accessible by the Luca Backend Server results in a wide range of potential harms the system could inflict on users, communities, and venues. In our analysis, we provide multiple examples of potential function creep - motivated by the high value the data generated by the Luca system might have for many entities. Amongst others, commercial actors could benefit from (real-time) information about venue occupancy and their past incidence numbers; law enforcement agencies could benefit from (real-time) information about the whereabouts of users to enforce safety policies or to conduct surveillance of target individuals and communities.

    \item The Luca system relies on a complex, centralised architecture in which the Luca Backend Server orchestrates and intermediates relationships between users, Health Departments, and Venue Owners. In the current design, the Luca Backend Server is the sole authority in the system that grants access to critical system functionalities and assigns roles to different entities. It can therefore not only observe all interactions, but also decide who has access to decryption keys, and who can request the decryption of records from different entities. Furthermore, the Luca service operator might change code and procedures at will without such changes being detected. This allows the Luca service operator to abuse and repurpose the data in the system without users, or even auditors, having the possibility to know that abuse is happening.
\end{itemize}

\vspace{2mm}
\noindent\textbf{Luca's security objectives}. The Luca Security Concept lists specific security objects.\footnote{See \url{https://luca-app.de/securityconcept/properties/objectives.html}} We argue below that these objectives are either achieved only under the assumption that the Luca Backend Server is trusted, or not met at all. We note that in our analysis we have identified many harms that are not covered by these security objectives.

    \begin{itemize}

        \item \emph{O1 ``An Uninfected Guest’s Contact Data is known only to their Guest App'':}\\
        A maliciously acting Luca server has several ways to decrypt the user references associated with check-ins (Sect. \ref{sec:active:venue} and \ref{sec:active:hd}), and therefore to gain access to users' contact data. This objective is therefore only achieved if the Luca Backend Server can be fully trusted and is neither compromised nor coerced to subvert the protections in place.

        \item \emph{O2 ``An Uninfected Guest’s Check-Ins cannot be associated to the Guest'':}\\
        Check-ins that belong to the same user are (partially) linkable (see Sect. \ref{sec:surveil:users}). This increases the risk that the corresponding Guests might be re-identified. Throughout our analysis, we show that the connections the Luca App uses to verify check-ins and perform check-outs can be linked to their IP address under normal operation and, if the server actively circumvents protection mechanisms, even to the user's \texttt{user\_id} (Sect. \ref{sec:surveil:users} and \ref{sec:stig:users}).
        
        \item \emph{O3: ``An Uninfected or Traced Guest’s Check-Ins cannot be associated to each other'':}\\ 
        As explained in the analysis, the Luca Backend Server has many means at its disposal to link check-in records, regardless of a Guests's status, often based solvely on data it observes during normal operation (see Sect. \ref{sec:surveil:users}). Moreover, through the normal tracing process, the Luca server obtains the \texttt{user\_id}s of all Traced Guests. Depending on the system's implementation and the number of traced contacts, this might enable the server to link check-ins by the same Traced Guest. As for O2, the objective is only achieved if the Luca Backend Server can be trusted.

        \item \emph{O4 ``An Infected Guest’s Check-In History is disclosed to the Health Department only after their consent'':}\\
        In our analysis, we have established that the Luca server can link check-ins to users (see O2) and link check-ins of the same user (see O3). Thus, the Luca server has the inference power to reveal any users' check-in history and share it with any entity, including the responsible Health Department, at any point in time. Thus, this objective holds only if the Luca server is trusted.

        \item \emph{O5 ``The Health Department learns only the relevant part of the Infected Guest’s Check-In History'':}\\
        A dishonest Luca Backend Server might attempt to reconstruct a user's check-in history at any point in time (see Sect. \ref{sec:surveil:users} and O4). As for O4, the objective is thus achieved only if the Luca server is trusted.

        \item \emph{O6 ``Traced Guest’s Contact Data is disclosed to the Health Department only after Venue Owners’ consent'':}\\
        A malicious or coerced Luca server has several ways to circumvent the protection provided by the venue owner encryption (Sect. \ref{sec:active:venue}. Therefore, the validity of this claim solely depends solely on the honesty of the Luca server.
    \end{itemize}

\vspace{2mm}
\noindent\textbf{Final remarks}. In conclusion, our analysis demonstrates how the deployment of digital presence tracing systems with centralised system architectures might dramatically increase the potential harms for individuals and communities. The centralised system design furthermore introduces new harms for venues with respect to their paper-based predecessors: venues need to be centrally registered and can be profiled in real time.

Our analysis raises the question whether the collection of detailed user information when implemented as a digital system can still be justified through its potential benefits. The trade-offs between the risks and benefits of sensitive data collection clearly shift when moving from a pure pen-and-paper-based system to a digital infrastructure that collects large amounts of  sensitive information about users and venues in a central location. Paper-based systems make it difficult for adversaries to exploit the collected information at a large scale, and do not imply any significant risks for venues. The introduction of digital, centralised, data-intensive solutions, however, considerably increase the potential harms of fine-grained data collection. The proportionality of data collection with respect to digital presence tracing solutions hence needs to be re-assessed. Furthermore, decentralized alternatives exist that can achieve similar, if not the same, functional goals but avoid large-scale data collection and minimise the risk for abuse. To make an informed decision about the deployment of digital presence tracing systems, these factors need to be taken into account.

\end{document}